\documentclass[pre,showpacs]{revtex4}
\usepackage{amssymb}
\usepackage[dvips]{graphicx}
\usepackage{amsmath}
\usepackage{graphicx}
\usepackage{makeidx}
\usepackage{subfigure}

\begin{document}

\title{Two-dimensional solitons and vortices in media with incommensurate
linear and nonlinear lattice potentials}
\author{Jianhua Zeng$^{1,2}$ and Boris A. Malomed$^{1}$}
\affiliation{$^{1}$Department of Physical Electronics, School of Electrical Engineering,
Faculty of Engineering, Tel Aviv University, Tel Aviv 69978, Israel\\
$^{2}$State Key Laboratory of Low Dimensional Quantum Physics, Department of
Physics, Tsinghua University, Beijing 100084, China}

\begin{abstract}
We construct families of ordinary and gap solitons (GSs), including
solitary vortices, in the two-dimensional (2D) system based on the
nonlinear-Schr\"{o}dinger/Gross-Pitaevskii equation with the 2D or
quasi-1D (Q1D) periodic linear potential, combined with the periodic
modulation of the cubic nonlinearity (also in the 2D or Q1D form),
which is, generally, incommensurate with the linear potential, thus
forming a ``nonlinear quasicrystal". Stable vortices are built as
complexes of four peaks with the separation between them equal to
the double period of the linear potential. The system may be
realized in photonic crystals or Bose-Einstein condensates (BECs).
The variational approximation (VA) is applied to ordinary solitons
(residing in the semi-infinite gap), and numerical methods are used
to construct solitons of all the types. Stability regions are
identified for soliton families in all the versions of the model.
\end{abstract}

\pacs{05.45.Yv, 03.75.Lm, 42.65.Tg}
\maketitle

\section{Introduction}

A versatile technique for the control of guided photonic and matter waves is
based on the use of periodic (lattice) potentials. In the case of
Bose-Einstein condensate (BEC) the potential may be induced by optical
lattices, and in photonics---by transverse structures in the form of
photonic crystals \cite{PC}. In nonlinear media, the lattice potentials help
to create and stabilize various types of solitons, both ordinary ones (found
in the semi-infinite gap of the underlying spectral structure \cite{Salerno}%
) and gap solitons (GSs) \cite{2Dgap}, which exist in the presence of the
self-attractive and repulsive nonlinearity, respectively. Along with the
fundamental solitons, the periodic potentials may support stable solitary
vortices, built as multi-peak patterns with the phase distribution carrying
the 
topological charge, in the case of the self-attraction \cite{Salerno} and
self-repulsion \cite{gap-vortex} alike.

Results accumulated in theoretical and experimental studies of one- and
two-dimensional (1D) and (2D) solitons (including 2D vortices) supported by
linear-lattice (LL) potentials were reviewed, respectively, in Refs. \cite%
{reviews-1D} and \cite{review-multiD}, see also a more recent review \cite%
{Barcelona-review}. A related topic is the study of discrete solitons in
optics, which correspond to the limit of very deep lattice potentials \cite%
{discrete-review}.

The photonic-crystal structures induce, simultaneously with the LL
potentials, an effective nonlinear potential (alias \textit{pseudopotential }%
\cite{Dong}), in the form of the concomitant spatially periodic modulation
of the local nonlinearity coefficient. In BEC, nonlinear lattices (NLs) may
be induced by external fields which affect the local nonlinearity via the
Feshbach resonance. The studies of solitons in NLs, as well as in LL-NL
combinations, have been recently reviewed in Ref. \cite{Barcelona}. A direct
experimental observation of NL-supported optical solitons was reported at a
surface between lattices \cite{experiment}.

A natural generalization of the setting combining the LL and NL is the one
with different or incommensurate periodicities of the two lattices. In the
framework of the 1D setting, both ordinary solitons and GSs, supported by
such a combination of competing linear and nonlinear potentials, were
studied in Ref. \cite{HS} by means of numerical methods and analytical
approximations. Noteworthy results were obtained, in particular, for
existence borders of the solitons as functions of the LL-NL
incommensurability, and for the empirical ``anti-Vakhitov-Kolokolov"
(anti-VK) stability criterion for GSs, which is written in terms of the
dependence of the chemical potential, $\mu $, on norm, $N$, of the soliton: $%
d\mu /dN>0$ (the VK criterion for ordinary solitons supported by the
self-attraction in the semi-infinite gap is $d\mu /dN<0$ \cite{Vakh}). The
objective of the present work is to produce results for 2D solitons
supported by incommensurate LL-NL combinations. We develop a variational
approximation (VA) which is applied, along with numerical methods, to
ordinary solitons, while GSs in the first finite bandgap are studied solely
in a numerical form, as the analytical approach would be too cumbersome in
that case. Numerical methods are also used to construct vortex solitons, in
the semi-infinite and finite gaps alike.

The settings considered here include both full 2D lattices potentials and
quasi-1D (Q1D) ones, which depend on the single coordinate (the LL potential
of the Q1D type is sufficient for the stabilization of ordinary 2D solitons,
in diverse realizations of 2D media with the self-attractive nonlinearity
\cite{BBB}). In fact, the combination of the periodic but mutually
incommensurate \ linear and nonlinear lattices makes the medium effectively
tantamount to a \textit{quasicrystal} for nonlinear excitations. Fundamental
solitons and solitary vortices in linear quasiperiodic potentials were
studied theoretically \cite{HS2}, and 2D photonic quasicrystals have been
recently created experimentally \cite{Q2D}. It is also relevant to mention a
recent work \cite{Abd}, which was dealing with 2D solitons in a model
combining crossed Q1D linear and nonlinear periodic potentials.

The rest of the paper is organized as follows. The model is introduced in
Section II, ordinary and gap solitons are considered, severally, in Sections
III and IV (each section reports the results for fundamental and vortex
solitons), and the work is concluded by Section V.

\section{The model}

The 2D system combining the periodic LL potential and NL
pseudopotential may be written in the form of the scaled
Gross-Pitaevskii (or nonlinear Schr\"{o}dinger) equation for the
BEC\ wave function (or the local amplitude of the electromagnetic
wave guided by the photonic crystal), $\psi \left( x,y,t\right) $
\cite{HS,Barcelona}:
\begin{eqnarray}
i\psi _{t} &=&-(1/2)\nabla ^{2}\psi -\varepsilon \left[ \cos (2\pi x)+\cos
(2\pi y)~\right] \psi  \notag \\
&&-g\left[ \cos (\pi qx)+\cos (\pi qy)\right] |\psi |^{2}\psi ,  \label{GPE}
\end{eqnarray}%
where $t$ is time (or the propagation distance in the photonic-crystal
waveguide), Laplacian $\nabla ^{2}=\partial _{x}^{2}+\partial _{y}^{2}$ acts
on transverse coordinates $x$ and $y$, the scale in the $\left( x,y\right) $
plane is set by fixing the LL period to be $1$, the period of the NL is $2/q$
(that is, $q$ is the \textit{incommensurability index}), and the NL strength
is normalized to be $g\equiv \pm 1$. The center of the soliton will be
placed at point $x=y=0$, hence $g=+1$ and $-1$ correspond, severally, to the
dominating self-attraction or self-repulsion, which support ordinary
solitons in the semi-infinite gap, or GSs in finite bandgap(s), respectively.

\begin{figure}[tbp]
\begin{center}
\includegraphics[height=4.cm]{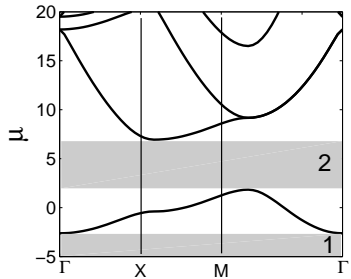}
\end{center}
\caption{The band structure of the linearized equation (\protect\ref{GPE})
in the first Brillouin zone, for depth $\protect\varepsilon =7.4$ of the
linear lattice; $\protect\mu $ is the chemical potential of the Bloch waves.}
\label{Fig1}
\end{figure}

The remaining parameter in Eq. (\ref{GPE}) is the normalized LL strength, $%
\varepsilon $. Generic results for the case when the system's spectrum
contains a single finite bandgap are reported here for a fixed strength, $%
\varepsilon =7.4$. The respective band structure in the first Brillouin zone
\cite{PC}, found from the linearized version of Eq. (\ref{GPE}), is
displayed in Fig. \ref{Fig1}.

At $q=0$, Eq. (\ref{GPE}) amounts to the usual model with the LL potential
and spatially uniform nonlinearity. The LL and NL are commensurate at $q=2$,
and the subharmonic commensurability, with the ratio of the LL and NL
periods $1:2$, occurs at $q=1$. The full incommensurability (overall
quasi-periodicity in the system) corresponds to irrational values of $q$,
but, in practical terms, the incommensurability may be emulated by $q=1.5$,
with the period ratio $3:4$. The Q1D versions of the model correspond to
dropping terms $\cos \left( 2\pi y\right) $ and/or $\cos \left( \pi
qy\right) $ in the linear and/or nonlinear potentials.

Stationary solutions to Eq. (\ref{GPE}) with chemical potential $\mu $ (or
propagation constant $-\mu $, in terms of the guided optical waves) are
looked for as $\psi (x,y,t)=\phi (x,y)\exp (-i\mu t)$, where function $\phi
(x,y)$ satisfies equation
\begin{eqnarray}
\mu \phi &=&-(1/2)\nabla ^{2}\phi -\varepsilon \left[ \cos (2\pi x)+\cos
(2\pi y)~\right] \phi  \notag \\
&&-g\left[ \cos (\pi qx)+\cos (\pi qy)\right] |\phi |^{2}\phi .  \label{phi}
\end{eqnarray}%
The VA will be based on the Lagrangian of Eq. (\ref{phi}), which is
\begin{eqnarray}
L &=&\int_{-\infty }^{+\infty }\left\{ \mu |\phi |^{2}-\frac{1}{2}|\nabla
\phi |^{2}+\varepsilon \left[ \cos (2\pi x)+\cos (2\pi y)\right] |\phi
|^{2}\right.  \notag \\
&&\left. +\frac{g}{2}\left[ \cos (\pi qx)+\cos (\pi qy)\right] |\phi
|^{4}\right\}dxdy .  \notag
\end{eqnarray}

\section{Localized modes in the semi-infinite gap}

\subsection{Fundamental solitons}

\begin{figure}[tbp]
\begin{center}
\includegraphics[height=4.cm]{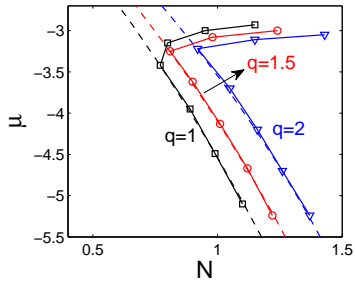}
\end{center}
\caption{(Color online) The chemical potential vs. the norm for soliton
families in the semi-infinite gap of the model with the full 2D linear and
nonlinear potentials, as produced by the VA (dashed curves), and as found
from numerical solutions of Eq. (\protect\ref{phi}) (chains of symbols).}
\label{Fig2}
\end{figure}

We start the analysis of the ordinary solitons, which are expected to exist
in the semi-infinite gap at $g=+1$ in Eq. (\ref{GPE}), using the VA based on
the Gaussian ansatz, $\phi \left( x,y\right) =A\exp \left[ -\left(
x^{2}+y^{2}\right) /\left( 2W^{2}\right) \right] $, with the corresponding
norm $N\equiv \int \int \phi ^{2}\left( x,y\right) dxdy=\pi \left( AW\right)
^{2}$ \cite{Salerno}. The substitution of the ansatz into the Lagrangian
yields the following expression, written in terms of $N$ and width $W$:%
\begin{equation}
L_{\mathrm{eff}}=\frac{N}{2}\left[ \mu -\frac{1}{2W^{2}}+2\varepsilon
e^{-\left( \pi W\right) ^{2}}+\frac{gN}{2\pi W^{2}}e^{-\left( \pi qW\right)
^{2}/8}\right] ,
\end{equation}%
and the respective variational equations, $\partial L_{\mathrm{eff}%
}/\partial W=\partial L_{\mathrm{eff}}/\partial N=0$:%
\begin{eqnarray}
\left( 2\pi W^{2}\right) ^{2}\varepsilon e^{-\left( \pi W\right) ^{2}}+\pi
^{-1}gNe^{-\left( \pi qW\right) ^{2}/8}\left[ 1+\left( \pi qW\right) ^{2}%
\right]  &=&1,  \notag \\
\left( 2W^{2}\right) ^{-1}-2\varepsilon e^{-\left( \pi W\right)
^{2}}-gN\left( \pi W^{2}\right) ^{-1}e^{-\left( \pi qW\right) ^{2}/8} &=&\mu
.  \label{VA}
\end{eqnarray}%
Dependences $\mu (N)$ for the soliton families, produced by a numerical
solution of Eqs. (\ref{VA}) at different values of incommensurability index $%
q$, are displayed in Fig. \ref{Fig2}, along with their counterparts,
produced by numerical solutions of stationary equation (\ref{phi}) and
verified in direct simulations of Eq. (\ref{GPE}).

\begin{figure}[tbp]
\begin{center}
\includegraphics[height=4.cm]{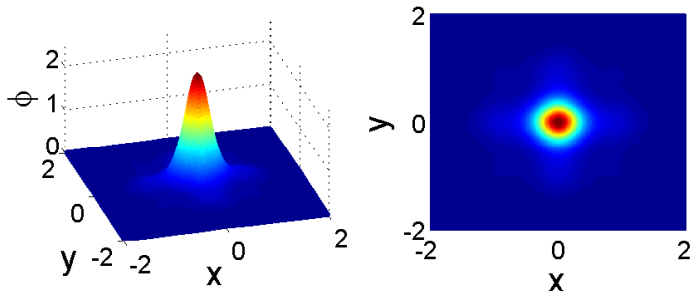}
\end{center}
\caption{(Color online) A typical example of a stable fundamental soliton
found in the semi-infinite gap of the model with the full 2D linear and
nonlinear potentials, for $g=+1$, $q=1.5$, and $\protect\mu =-5.24,$ $N=1.22$%
. The right panel shows contour plots of the stationary real wave function, $%
\protect\phi \left( x,y\right) $, for the soliton.}
\label{Fig3}
\end{figure}

\begin{figure}[tbp]
\begin{center}
\includegraphics[height=4.cm]{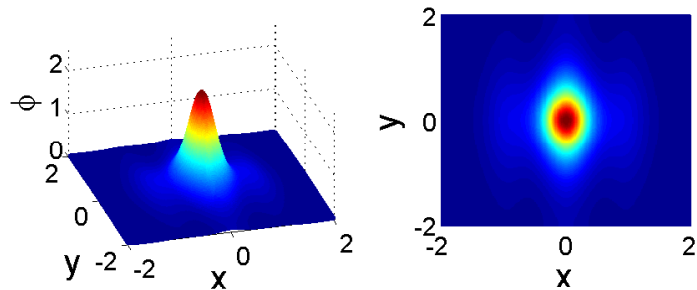}
\end{center}
\caption{(Color online) The same as in Fig. \protect\ref{Fig3}, but in the
case of the Q1D linear potential (while the nonlinear potential remains
two-dimensional), for $g=+1$, $q=1$, and $\protect\mu =-2.91,$ $N=1.68$. The
shape of the solitons in the version of the model where the nonlinear
potential is also made quasi-one-dimensional is quite similar \ to the
present one. }
\label{Fig4}
\end{figure}

Shapes of the stable solitons generated in the semi-infinite gap by the full
2D model, and by its version with the Q1D linear potential, are displayed in
Figs. \ref{Fig3} and \ref{Fig4}. Naturally, these shapes are, respectively,
quasi-isotropic and strongly elongated, resembling those reported previously
in other 2D models stabilized by the LL potentials \cite%
{Salerno,review-multiD,Barcelona,Abd}.

It is seen from Fig. \ref{Fig2} that the VA describes the ordinary solitons
with a reasonable accuracy, except near the edge of the semi-infinite gap,
where the Gaussian ansatz is irrelevant, as the soliton becomes very wide
and develops a complex shape. Further, simulations of the evolution of
perturbed solitons demonstrate that the stability of the solitons exactly
obeys the VK criterion, $d\mu /dN<0$ [strictly speaking, if dependence $\mu
(N)$ is taken in the numerically found form]. These features of the families
of ordinary-soliton solutions are similar to those found before in the 1D
variant of the model \cite{HS}.

\begin{figure}[tbp]
\begin{center}
\includegraphics[height=4cm]{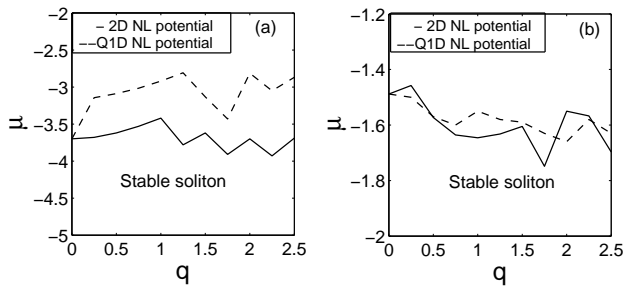}
\end{center}
\caption{(a) Stability borders for the family of the ordinary solitons (in
the semi-infinite gap) in the model with the 2D linear potential and 2D or
Q1D nonlinear one (solid and dashed lines, respectively). (b) The same, in
the case of the Q1D linear potential. The solitons are stable beneath the
respective borders.}
\label{Fig5}
\end{figure}

The stability of the 2D solitons in the semi-infinite gap, for all the four
realizations of the model (2D or Q1D linear and nonlinear potentials) is
summarized in Fig. \ref{Fig5}. When solitons are unstable, they suffer decay
into radiation waves (rather than rearranging themselves into stable
solitons). It is worthy to note that the replacement of the full 2D NL by
its Q1D counterpart leads to an expansion of the stability areas, which is
explained by the fact that, in the case of the Q1D NL, the ordinary solitons
should make the effort to ``dodge" the destabilizing locally self-repulsive
nonlinearity only in one direction, rather than in two.

In addition to the above analysis, we tried to test the mobility of
the solitons by simulating their evolution after sudden application
of a ``kick", i.e., multiplication of the wave function of a stable
quiescent soliton by the phase-tilt factor, $\exp \left( i\left(
k_{x}x+k_{y}y\right) \right) $, with vectorial kick parameter
$\mathbf{k}$. Except for the obvious case when both the LL and NL
have the collinear Q1D structure, and the kick is applied in the
unconfined direction, mobile solitons were not found, even if either
the LL or NL potential was of the
Q1D type.
Instead of setting the soliton in motion, a sufficiently strong kick tends
to destroy it.

\subsection{Solitary vortices}

\begin{figure}[tbp]
\begin{center}
\includegraphics[height=7.cm]{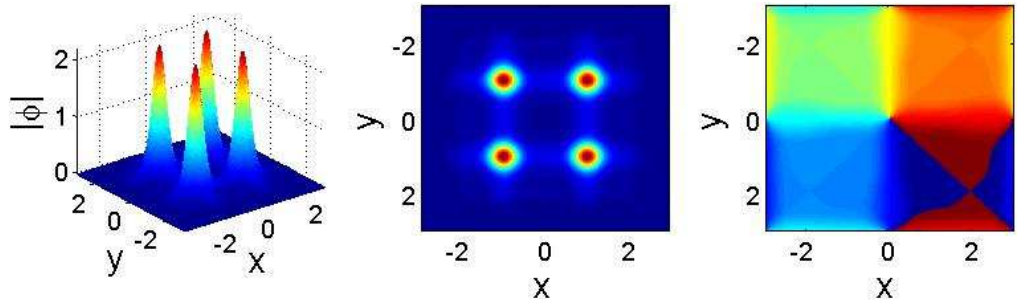}
\end{center}
\caption{(Color online) An example of a stable vortex with topological
charge $1$, which is supported, in the semi-infinite gap, by the combination
of the full 2D linear and nonlinear structures with $q=1$ and $g=+1$. The
left, middle, and right panels display, respectively, the absolute value of
the field, $|\protect\phi (x,y)|$, as a function of the coordinates, contour
plots of $|\protect\phi (x,y)|$, and the distribution of the phase, which
carries the vorticity. Parameters of the vortex are $\protect\mu =-5.1$, $%
N=4.4$.}
\label{Fig6}
\end{figure}

Stable solitary vortices with topological charge $1$ were found as
``hollow" four-peak complexes, with the separation between the peaks
equal to the double period of the LL potential ($\Delta x=\Delta
y=2$), and a nearly empty cell at the center of the pattern, see an
example in Fig. \ref{Fig6}. It is relatively easy to find stable
vortices of this type, due to the weak interaction between the
peaks. More densely packed vortex patterns can also be constructed,
but we could not find stability regions for them. It is known from
the studies of other models too that the vortices with inner
``voids" are more likely to be stable
\cite{review-multiD,Barcelona}.

Actually, the vortices are stable only for values of the incommensurability
index close to $q=0,~1,$ and $2$---namely, within intervals of half-width $%
\Delta q\simeq 0.1$ around these values. The latter observation may be
explained by the fact that, at such values of $q$, both the linear and
nonlinear potentials have minima at or close to sites where the the power
(density) peaks are located.

\begin{figure}[tbp]
\begin{center}
\includegraphics[height=4.cm]{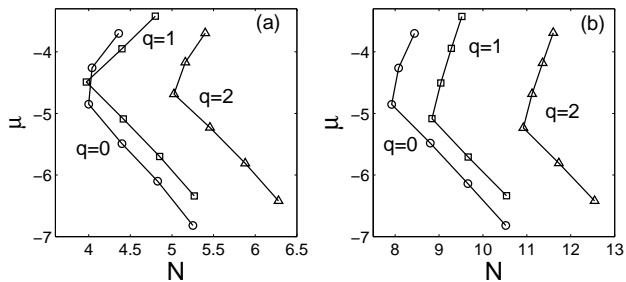}
\end{center}
\caption{(Color online) The same as in Fig.2, but for families of solitary
vortices in the semi-infinite gap, supported by the combination of the full
2D linear potential with the 2D (a) or Q1D (b) nonlinear lattice.}
\label{Fig7}
\end{figure}

Furthermore, the vortices of the type shown in Fig. \ref{Fig6} are found to
be stable (in the semi-infinite gap) in the case when the LL potential is
fully two-dimensional, while the NL may be of either 2D or Q1D type. The
corresponding families of the vortex modes are represented by $\mu (N)$
curves in Fig. \ref{Fig7}. Detailed analysis demonstrates that the stability
of these families exactly follows the VK criterion, i.e., stable are
portions of the families with $d\mu /dN<0$.

\section{Gap solitons and vortices}

\subsection{Fundamental solitons}

\begin{figure}[tbp]
\begin{center}
\includegraphics[height=4.cm]{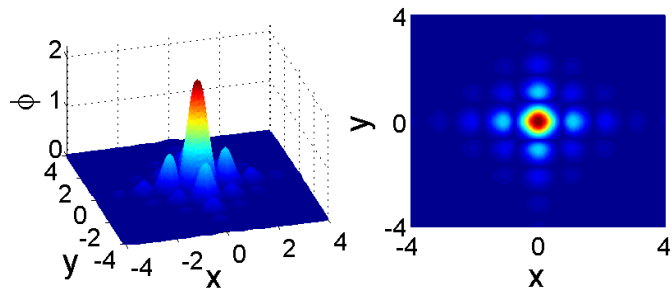}
\end{center}
\caption{(Color online) A typical example of a stable gap soliton in the
model with the full 2D NL, for $g=-1,q=1.5,$ and $\protect\mu =3.4,N=1.4$.}
\label{Fig8}
\end{figure}

\begin{figure}[tbp]
\begin{center}
\includegraphics[height=4.cm]{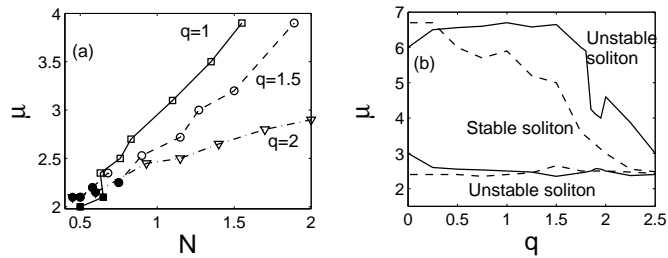}
\end{center}
\caption{(a) Curves $\protect\mu (N)$ for families of gap solitons in the
model with the full 2D nonlinear potential. Black squares denote unstable
solitons close to the bottom edge of the bandgap. (b) Stability borders for
the entire set of the GS families in the model with the 2D and Q1D nonlinear
potentials (solid and dashed lines, respectively). The solitons are stable
between the corresponding stability borders. Both the top and bottom borders
are located near the respective edges of the finite bandgap.}
\label{Fig9}
\end{figure}

Numerically generated fundamental GSs feature, as usual, more complex shapes
than the ordinary solitons, see an example of a stable GS, found
sufficiently far from edges of the finite bandgap, in Fig. \ref{Fig8} (in
the model combining the 2D LL and Q1D NL, the shapes of the GSs are quite
similar). The GS families are characterized by $\mu (N)$ curves which are
shown in Fig. \ref{Fig9}(a). Unlike the ordinary solitons (cf. Fig. \ref%
{Fig2}), the GSs always feature $d\mu /dN>0$, thus complying with
the ``anti-VK" criterion proposed in Ref. \cite{HS}.

Numerical tests demonstrate that the GSs tend to be stable sufficiently deep
inside the finite bandgap, and unstable near its edges (unstable GSs suffer
decay into radiation). The numerically found stability borders for the
entire set of GSs in the two versions of the present model, with the full 2D
NL and its Q1D counterpart, are presented in Fig. \ref{Fig9}(b). It is
observed that, on the contrary to the solitons in the semi-infinite gap, the
stability region for GSs tends to be essentially narrower for the NL of the
Q1D type, in comparison with the full 2D NL. The latter feature seems
natural, as, unlike the case of ordinary solitons, the NL may provide for an
additional support to the GSs.

With the increase of $q$, the GS\ stability areas clearly tend to shrink to
nil, which actually happens in Fig. \ref{Fig9}(b) for the variant of the
model with the Q1D NL (we expect the same ought to happen for the full 2D
NL, but numerical problems impede extending the stability diagram to still
larger values of $q$). This trend is explained by the fact that, at large $q$%
, the rapidly oscillating NL field tends to average itself to zero, hence
the broad (see Fig. \ref{Fig8}) GS ceases to feel the action of the
nonlinearity. The ordinary solitons in the semi-infinite gap do not
demonstrate such a trend (cf. Fig. \ref{Fig5}), as, following the increase
of $q$, these solitons are able to compress themselves inside a single cell
of the structure, remaining centered around a region with the
self-attractive nonlinearity. Finally, numerical tests demonstrate that, as
well as it was concluded above for the ordinary solitons, the GSs are not
mobile objects (not shown here in detail).

\subsection{Solitary vortices}

\begin{figure}[tbp]
\begin{center}
\includegraphics[height=7.cm]{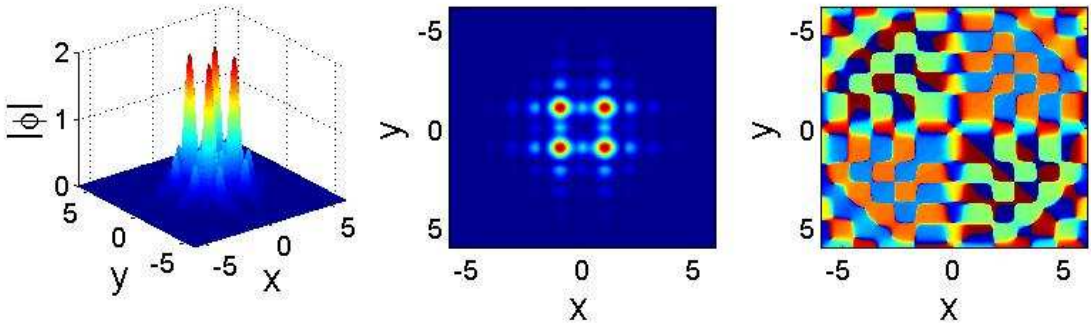}
\end{center}
\caption{(Color online) An example of the stable vortex soliton found in the
finite bandgap, for $q=1,$ $g=-1$ and $\protect\mu =3.5,$~$N=5.36$. The
meaning of the three panels is the same as in Fig. \protect\ref{Fig6}, }
\label{Fig10}
\end{figure}

Stable vortex solitons that can be found in the finite bandgap feature the
same structure which supports the stable vortices in the semi-infinite gap
(cf. Fig. \ref{Fig6}), i.e., they are built of four peaks separated by the
distance equal to the double LL period, the vorticity being carried by the
corresponding phase distribution, see an example in Fig. \ref{Fig10}. The
difference from the situation in the semi-infinite gap is that the solitary
vortices in the finite bandgap may be stable only when \emph{both} LL and NL
have the full 2D structure (i.e., the vortices are unstable if the NL is of
the Q1D type).

\begin{figure}[tbp]
\begin{center}
\includegraphics[height=4.cm]{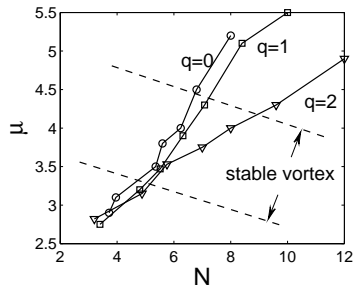}
\end{center}
\caption{The dependence of the chemical potential on the norm for families
of solitary vortices in the finite bandgap. Stable portions of the vortex
families are located inside the marked stripe.}
\label{Fig11}
\end{figure}

As well as their counterparts in the semi-infinite gap, the solitary
vortices in the finite bandgap are found to be stable only for the
values of incommensurability index close to $q=0,$ $1,$ and $2$.
Families of these vortices are represented in Fig. \ref{Fig11} by
the corresponding $\mu (N)$ curves. As well as the fundamental gap
solitons, the stable vortices in the finite gap obey the ``anti-VK
criterion, $d\mu /dN>0$.

\section{Conclusions}

We have introduced the model of 2D nonlinear photonic crystals and
BEC based on the interplay of linear and nonlinear lattices with
different (generally, incommensurate) periods, which may be
considered as a ``nonlinear quasicrystal". Both fully 2D periodic
potentials and their Q1D reductions were considered. For the
ordinary solitons in the semi-infinite gap, the VA\ (variational
approximation) was developed. In the general case, the solitons and
solitary vortices were explored by means of numerical methods. The
stability regions have been identified for the entire sets of the
ordinary solitons and GSs (gap solitons). Stable families of vortex
solitons of both types have been found too.

A relevant direction for the development of the analysis reported above may
be search for stable higher-order vortices. 
On the other hand, it may also be interesting to extend the analysis to a
broader parameter region, which may give rise to higher-order bandgaps, in
addition to the single finite bandgap existing in the situation considered
in this work, and construct solitons and solitary vortices in the additional
gaps.

The work of J.Z. was supported by a postdoctoral fellowship provided by the
Tel Aviv University, and by grant No. 149/2006 from the German-Israel
Foundation.


\begin{thebibliography}{99}
\bibitem{PC} Joannopoulos J D , Johnson S G, Winn J N and Meade R D 2008
\textit{Photonic Crystals: Molding the Flow of Light }(Princeton University
Press: Princeton); Skorobogatiy M and Yang J 2009 \textit{Fundamentals of
Photonic Crystals Guiding} (Cambridge University Press: Cambridge).

\bibitem{Salerno} Baizakov B B, Malomed B A and Salerno M 2003 \textit{%
Europhys. Lett.} \textbf{63} 642-8; Yang J and Musslimani Z H 2003 \textit{%
Opt. Lett.} \textbf{28} 2094-6.

\bibitem{2Dgap} Baizakov B B, Konotop V V and Salerno M 2002 \textit{J.
Phys. B: At. Mol. Opt. Phys.} \textbf{35} 5105-19; Ostrovskaya E A and
Kivshar Y S 2004 \textit{Opt. Exp.} \textbf{12} 19-29.

\bibitem{gap-vortex} Sakaguchi H and Malomed B A 2004 \textit{J. Phys. B:
At. Mol. Opt. Phys.} \textbf{37} 2225-39; Ostrovskaya A E and Kivshar Y S
2004 \textit{Phys. Rev. Lett.} \textbf{93} 160405-4.


\bibitem{reviews-1D} Brazhnyi V A and Konotop V V 2004 \textit{Mod. Phys.
Lett. B} \textbf{18} 627-51; Morsch O and Oberthaler M 2006 \textit{Rev.
Mod. Phys.} \textbf{78} 179-215.

\bibitem{review-multiD} Malomed B A, Mihalache D, Wise F and Torner L 2005
\textit{J. Optics B: Quant. Semicl. Opt.} \textbf{7} R53-R72.

\bibitem{Barcelona-review} Kartashov Y V, Vysloukh V A, and Torner L 2009
\textit{Progress in Optics} \textbf{52} 63-148 (ed. by E. Wolf: North
Holland, Amsterdam).

\bibitem{discrete-review} Lederer F, Stegeman G I, Christodoulides D N,
Assanto G, Segev M and Silberberg Y 2008 \textit{Phys. Rep.} \textbf{463}
1-126.

\bibitem{Dong} Mayteevarunyoo T, Malomed B A and Dong G 2008 \textit{Phys.
Rev. A} \textbf{78} 053601-12 .

\bibitem{Barcelona} Kartashov Y V, Malomed B A, and Torner L 2011 \textit{%
Rev. Mod. Phys.} \textbf{83} 247-305.

\bibitem{experiment} Kartashov Y V, Vysloukh V A, Szameit A, Dreisow F,
Heinrich M, Nolte S, T\"{u}nnermann A, Pertsch T and Torner L 2008 \textit{%
Opt. Lett.} \textbf{33} 1120-2.

\bibitem{HS} Sakaguchi H and Malomed B A 2010 \textit{Phys. Rev. A} \textbf{%
81} 013624-9.

\bibitem{Vakh} Vakhitov M and Kolokolov A 1973 \textit{Izvestiya VUZov
Radiofizika} \textbf{16} 1020-8 [English translation: 1973 \textit{%
Radiophys. Quantum. Electron.} \textbf{16} 783-9]; Berg\'{e} L 1998 \textit{%
Phys. Rep.} \textbf{303} 259-370.

\bibitem{BBB} Baizakov B B, Malomed B A and Salerno M 2004 \textit{Phys.
Rev. A} \textbf{70} 053613-9; Mihalache D, Mazilu D, Lederer F, Kartashov Y
V, Crasovan L C and Torner L 2004 \textit{Phys. Rev. E} \textbf{70}
055603(R)-4; Gubeskys L and Malomed B A 2009 \textit{Phys. Rev. A} \textbf{79%
} 045801-4.


\bibitem{HS2} Sakaguchi H and Malomed B A 2006 \textit{Phys. Rev. E} \textbf{%
74} 026601-7.

\bibitem{Q2D} Levi L, Rechtsman M, Freedman B, Schwartz T, Manela O and
Segev M 2011 \textit{Science} \textbf{332} 1541-4; Vardeny Z V and Nahata A
2011 \textit{Nature Photonics} \textbf{5} 453-3.

\bibitem{Abd} da Luz H L F, Abdullaev F K, Gammal A, Salerno M and Tomio L
2010 \textit{Phys. Rev. A} \textbf{82} 043618-8.
\end{thebibliography}
\end{document}